# Role of Mentorship, Career Conceptualization, and Leadership in Developing Women's Physics Identity and Belonging

Jessica L. Rosenberg[1*], Nancy Holincheck[2], Kathryn Fernández[1], Benjamin W. Dreyfus[1,3], Fardousa Wardere[2], Stephanie Stehle[2], Tiffany N. Butler[2]
[1]Department of Physics and Astronomy, George Mason University, Fairfax, VA 22030
[2]College of Education and Human Development, George Mason University, Fairfax, VA 22030
[3]STEM Accelerator Program, College of Science, George Mason University, Fairfax, VA 22030
* Correspondence: jrosenb4@gmu.edu

## ABSTRACT

The percentage of women receiving bachelor's degrees in physics (25%) in the U.S. lags well behind that of men, and women leave the major at higher rates. Achieving equity in physics will mean that women stay in physics at the same rates as men, but this will require changes in the culture and support structures. A strong sense of belonging can lead to higher retention rates so interventions meant to increase dimensions of physics identity (interest, recognition, performance, and competence) may increase persistence overall and increase women's retention differentially. We describe our model in which mentorship, an understanding of career options (career conceptualization), and leadership are inputs into the development of these dimensions of physics identity. This paper includes preliminary results from a qualitative study that aims to better understand how career conceptualization, leadership, and mentorship contribute to the development of physics identity and belonging. We report results from a survey of 15 undergraduate physics students which was followed up by interviews with 5 of those students. The students were from 2 institutions: a small private liberal arts college in the Midwest region of the U.S. and a large public university in the southeast region of the U.S. classified as a Hispanic-serving institution (HSI). With respect to mentorship, consistent with the existing literature, we found that it could provide critical support for students' engagement in the physics community. Leadership experiences have not previously been positioned as an important input into identity, yet we found that they helped women in physics feel more confident, contributing to their recognition of themselves as physics people. While the data on how career conceptualization contributed to the building of identity is limited, there are some connections to recognition and competence, and it will be an interesting avenue of future exploration.

## I. INTRODUCTION

Although some STEM fields have moved toward gender parity, physics remains markedly unbalanced. Achieving parity will require that women stay in physics at the same rates as men, but this, in turn, requires effective support structures that promote identity and belonging. The workforce growth in STEM continues to outpace overall workforce growth [1], yet in the U.S., women still make up only 29% of physical scientists (they are 52% of the college-educated workforce) and earn only 25% of the bachelor's degrees in physics [2]. Increasing the proportion of women and people from underrepresented groups in physics is important for the strength of physics as a discipline and for the growth of the science and engineering workforce.

To strengthen and grow physics as a discipline, students will need to be prepared for a broad array of STEM careers. The Joint Task Force on Undergraduate Physics Programs (J-TUPP) Phys21 report [3] pointed out that 95% of physics bachelor's degree recipients look for employment outside of academia. The structure and environment of physics disciplines have not always made for an inclusive space, and women have tended to struggle more than men with building a sense of physics identity and belonging [4], so interventions meant to increase dimensions of physics identity (interest, recognition, performance, and competence) [5] may increase their persistence in STEM and positively influence their career choice.

In this exploratory study, we took a qualitative approach to understanding how undergraduate students build their sense of physics identity and belonging through mentorship, professional development with a focus on understanding career pathways, and leadership activities. Our theoretical model extends a well-established model of physics identity developed in prior qualitative [6] and quantitative [5] research. Our study focuses on how specific theorized factors may have influenced our participants' physics identity formation. By examining these dimensions, we hope to help support the creation of effective mentorship, leadership, and professional development programs for undergraduate physics students.

## II. MODEL OF PHYSICS IDENTITY AND RESEARCH QUESTIONS

In our exploration of physics identity, we draw on Gee's [7] definition of identity as recognizing oneself and being recognized by others as a certain kind of person. Gee [8] noted an aspirational aspect of identity in that it represents the kind of person one wants to be in the future, as well as in the present. Wenger [9] described identity in terms of "learning trajectory," stating that "we define who we are by where we have been and where we are going." Gee's conception of identity as aspirational is both related to and distinct from Wenger's description of identity as being shaped by our past, present, and future experiences. The malleability of identity over time is of particular importance in our project, as we consider identity through the experiences of women in physics and their conceptualization of future career options.

### A. Background of Science and Physics Identity

Science identity is best understood through the foundational work of Carlone and Johnson [6], who developed a model of science identity with three interrelated dimensions: (1) performance of scientific practices, (2) competence as knowledge and understanding of content, and (3) recognition of oneself and by others as a science person. Hazari et al. [5] developed a model of physics student identity built on that of Carlone and Johnson and identified interest as an additional dimension that influences students' decisions of who and what they want to be. Hazari and colleagues found statistically significant correlations between performance, competence, recognition, and interest and "seeing oneself as a physics person." The Hazari model describes these four components of identity as key to students' perception of themselves, framing performance as "belief in the ability to perform required physics tasks" and competence as "belief in the ability to understand physics content." This model has been widely applied to understand the discipline-specific identity of students across STEM, including engineering, mathematics, and science [10–15].

The idea that the sense of self is at the heart of identity is echoed by Kane [16], who talks about how, for children, the importance of each of these dimensions can vary in forming science identity. They also note that the perception of performance, competence, and recognition can differ. Beyond differing perceptions of these dimensions by different students, Hazari et al. [5] have looked at how a sense of identity can change as undergraduates progress through their program of study and find themselves in different environments, surrounded by different groups of people. The personal and fluid nature of identity emphasizes the need for this to be a self-identified concept over a "looser" definition that allows for the ability to observe someone else's sense of science identity (Danielson et al., [17]). All of the key aspects of identity are interpreted by the individual and are subject to their understanding of these ideas and of their environment.

Students with a strong science identity have a higher likelihood of choosing a science career [18] and are more likely to demonstrate persistence in STEM courses and careers [6]. Multiple studies [4,19] have connected identity and persistence in STEM and found gender differences in science identity. Research on student persistence in STEM has established that students who choose to leave STEM majors and careers are as competent in STEM as their peers [20,21]. A lack of persistence in STEM is often not related to student ability but to the culture of undergraduate science departments and classrooms failing to cultivate science identity in some students [22]. In recent years, the connection between STEM capital and STEM identity has demonstrated that a variety of experiences can contribute to students' STEM identity [23]. Potvin et al. [24] also showed the importance of direct intervention through exposure to counternarratives in increasing interest in pursuing physics for students who identify as female.

The literature discussed here draws from efforts to understand both science and physics identity. It is important to note that while there is a strong correlation between physics identity and science identity, physics has its own culture and norms that are distinct from science as a whole and are relevant for students pursuing a physics degree.

### B. Theoretical Inputs to Physics Identity

Our model (see Fig. 1) builds upon the identity framework of Hazari et al. [5], which describes a physics student's identity as consisting of personal, social, and physics identities. Physics identity in the framework is composed of performance, recognition, interest, and competence with respect to physics. Performance, recognition, interest, and competence are, in turn, built by student experiences and the contextualization of those experiences. For the purposes of this work, we examine how mentorship and leadership experiences and students' understanding of career opportunities are inputs into, and thus build, performance, recognition, interest, and competence, which in turn build physics identity.

Our model places *self-efficacy* at the intersection of performance and competence. Bandura defines *self-efficacy* as an individual's belief that they can be successful when *performing* a task in a specific context and ties it to the persistence displayed when working towards related goals [25]. This performance, in turn, can impact a student's view of their competence [26]. Self-efficacy is both related to and distinguished from an individual's outcome expectancy beliefs, or the consequences expected from one's own actions [27]. In Carlone and Johnson's [6] model of science identity, an individual's belief about their own *competence* in science (e.g. 'I understand science') and their belief about their ability to demonstrate this *competence* through *performance* of science (e.g. 'I can show that I understand science in tests, lab activities or discussions') were intertwined with *how one is recognized* as a science person by oneself and others. An individual's sense of *competence* in science contributes to their sense of *self-efficacy* in science, or confidence in *ability to perform* scientific tasks. Research in STEM education has tied self-efficacy to academic persistence [28,29] and STEM career interest [30], and has also identified gender- and race-based differences in student self-efficacy [31,32].

Hazari et al. [25] define a *sense of belonging* as a student's perception that they fit in or do not feel excluded within their physics community. Women's sense of belonging has been tied to their grade in a physics course [33] and to their intention to persist in physics and calculus [34]. In a factor analysis study, Kalendar [35] found that women's perception of being *recognized* as a physics person by their teaching assistant was related to women's *sense of belonging* and *self-efficacy* in physics. Students' *sense of belonging* within the physics community is connected to their *intersectional identities*, especially social, physics, and personal identities, which is why we place it at the intersection of the identities in Fig. 1. Avraamidou [36] examined the identities of minoritized women in physics and found that the impact of students' social and personal identities hindered the students' sense of belonging in physics. This study also demonstrated the role other women in physics, including a teaching assistant, played in helping participants feel some sense of belonging in physics.

This model illustrates that the development of any one component of a student's identity (e.g., their physics identity) is dependent on the connections between that identity and the other components of their intersectional identity.

Our initial theorization of how leadership contributes to women's physics identities was based upon our own experiences as well as some reading on leadership identity more generally. Six of the seven authors of the study are women in STEM, and four of us are women in physics. As we reviewed the literature and considered how undergraduate women might develop physics identities, we noted that students' formal and informal leadership experiences in physics and STEM, including in physics clubs, are a factor that had not been considered. We hypothesized that leadership activities could contribute to *recognition*, *performance*, and *competence* but that they could also be their own important dimension of identity, as both *leadership self-efficacy* and *leadership identity* have been discussed in the literature (e.g., [37–40]). Since the leadership literature is not focused on undergraduate women in physics, it has not explored whether it is the ability to identify as a leader that helps to build physics identity or the contribution of leadership experiences to feelings of recognition, performance, and competence. While this preliminary study will not allow us to distinguish these possibilities, we hope to better understand where leadership belongs in the model as the project progresses.

There is a gendered component of leadership identity that may be important for undergraduate women in STEM fields. An examination of how undergraduate women in computing define leadership showed that they tended to describe the importance of relationships and interpersonal skills. However, when they were then asked to assess their own leadership skills, they focused on their "perceived inability to be authoritative of their concern about being under scrutiny in the public eye" [41]. A study of undergraduate learning assistants, physics students focused on helping peers in classroom settings, found that leadership identity contributes to a student's physics identity through the building of competence [42]. Other work has shown that leadership experiences and leadership identity are deeply entwined and built recursively through social processes that can have positive and negative feedback loops [43,44]. Positive feedback loops lead to increased self-confidence and motivation to lead [45,46]. We seek to look more broadly at students' physics leadership experiences, including both formal and informal leadership roles, and the ways that they impact students' sense of identity through building a sense of competence and feelings of recognition, but also how their identity as a leader impacts their sense of identity as a physics person.

In our model, mentorship plays a role in helping students build a physics identity. We explore the ways in which faculty and *peer mentorship* promote *recognition* and influence *identity*. Effective mentorship can take different forms and often requires multiple mentors who provide support for students in developing physics identity [47,48]. Peer mentors can help inoculate students against negative stereotypes and other negative experiences that impact their *self-efficacy* and *sense of belonging* [49,50]. Faculty and peers can play an important role as mentors by helping to build *recognition, competence*, and, possibly, *interest* [51]. Mentoring relationships have been found to be impactful for both mentors and mentees in STEM settings [52]. We examine how different mentors influence students' development of their physics identity. We note that while we were also interested in the impact of professional mentors, the students in this preliminary sample did not have experience with that kind of mentorship, so we do not discuss it here.

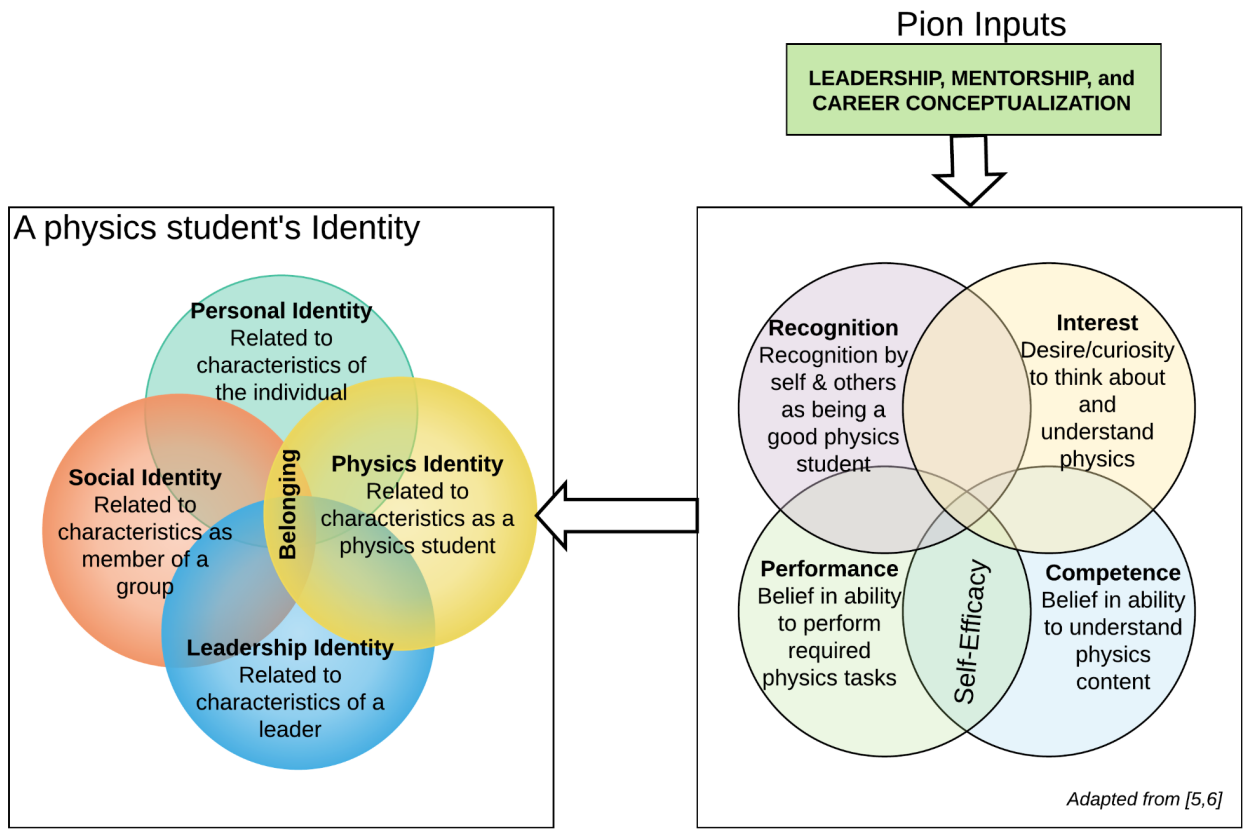

FIG. 1. Our theoretical model builds on the identity framework of Hazari et al. [5], which describes a physics student's identity as consisting of personal identity, social identity, and physics identity (left box). Recognition, performance, interest, and competence are important for students' ability to build physics identity [5], as shown in this diagram with an arrow pointing at the physics identity circle. We theorize that leadership and mentorship experiences and career conceptualization can serve as inputs to support the development of physics identity for undergraduate women. Our model places self-efficacy at the intersection of performance and competence, and this performance, in turn, demonstrates their competence [26]. Students' sense of belonging within the physics community [56] occurs at the intersection of the elements of their identity and is particularly related to social, physics, and personal identities. Note that it is not yet clear whether leadership is significant as an independent component of a student's identity or as an input into recognition, performance, and competence (it may not impact interest directly) to build a student's physics identity.

Interest is a key element in the development of identity [5], but we hypothesize that translating an interest in physics into an identity as a physicist may be particularly tricky for students who do not intend to pursue an academic path and are unsure of their career options beyond their undergraduate degree. Understanding the broad range of careers that they can access with a physics degree may help students who are not interested in an academic career path identify as physicists. This is particularly relevant because non-academic careers are the most common path, with 52% of physics bachelor's degree recipients (2019 and 2020) in the workforce a year after graduation [53]. Some research has examined why physics Ph.D. students are interested in non-academic careers (e.g., [54,55]), but no researchers have examined the physics identity of these students. We examine how students' *career interests* and knowledge of career options influence their physics interests and the development of *physics identity*.

### C. Research Questions

1. How does the development of faculty and peer mentoring relationships impact the development of a sense of belonging and physics identity for women in physics?
2. How does understanding the career options (career conceptualization) for a physicist impact undergraduate women's sense of belonging and physics identity?
3. How does the development of leadership identity increase a sense of belonging and physics identity for women in physics?

### III. METHODS

This study reports the preliminary findings from a grant-funded project focused on the experiences of women enrolled in undergraduate physics programs. During the first year of the project, our team collected survey and interview data from 15 physics students at two institutions in the United States. Both institutions were identified for participation in the study due to the robust number of bachelor's of physics graduates and the presence of an active chapter of the Society of Physics Students (SPS, a national organization of college/university physics clubs) or Sigma Pi Sigma (an undergraduate physics honor society connected with SPS). We focused on institutions with large and active SPS and Sigma Pi Sigma chapters because these organizations often sponsor mentorship and professional development opportunities while providing leadership opportunities for their club officers. One of the institutions is a small private liberal arts college in the Midwest region of the U.S., and the other is a large public university in the southeast region of the U.S. classified as a Hispanic-serving institution (HSI). In our recruitment, the project team contacted faculty or students at these schools through our professional networks and identified a physics major to serve as the project liaison. The liaison promoted our project to their peers, distributed links to surveys to all members of the SPS and Sigma Pi Sigma chapters and met with the research team several times to share information about the department and SPS mentorship and professional development activities that supported physics students' understanding of career options.

### A. Participants and Data Collection

Fifteen undergraduate students at these two US institutions completed an initial survey, including eight students who identified as men and seven who identified as women (the option to identify as non-binary or self-defined gender was available but not taken by respondents). Survey participants included nine seniors, five juniors, one sophomore, and three students who did not designate their year in school. All students were physics majors; eight had a double major in another STEM discipline (i.e., mathematics, biology, chemistry, data science), and six identified their minor as astronomy, education, or mathematics. The survey group included six Hispanic students, one Asian student, and eight White students. The study was approved by the institutional research review board, and all participants consented to participate in the research.

The survey served multiple purposes, including identifying potential interview participants and collecting initial data about students' experiences in physics. We included forced-response and short-answer questions about students' interests in workshops, mentoring, and physics club social activities to help the liaison identify activities that could be planned. To encourage quality responses [57], we included only one long-form open-ended question, "What has your experience been like as a physics student?" Participants provided one to three robust paragraphs in their responses to this question.

We invited all women who completed the survey to participate in a one-hour interview about their experiences in physics. The interviews used a semi-structured protocol aimed at understanding the students' experience within physics. The interviews were broken up into three sections with questions about (1) the student's story, experiences, and future plans, (2) mentorship and professional development programs in which the student had participated, and more in-depth questions about mentorship and leadership experiences in physics, STEM, or elsewhere, and (3) questions about whether they have felt valued in their physics department and challenges or barriers that they have overcome.

Five women from 2 institutions agreed to be interviewed, including both student liaisons (see Table 1 which includes pseudonyms for the participants, their race/ethnicity, year in school, and major). Note that the institution attended by Gabriella and Nina is a Hispanic-serving institution, while Goldie, Daria, and Violet all attend a small liberal arts college in the Midwest. The interviews were conducted via Zoom and were transcribed and analyzed by members of the research team.

The long-form survey responses provided some of the context for this work, but most of our results have been drawn from the interviews of the five students listed in Table 1.

TABLE I. Interview participant information

| Participant | Race/Ethnicity | Year | Major |
| --- | --- | --- | --- |
| Gabriella | Hispanic/Latina | Junior (3rd year) | Physics |
| Nina | Hispanic/Latina | Senior (4th year) | Physics |
| Daria | White | Senior (4th year) | Physics/Math |
| Violet | White | Senior (4th year) | Physics/Psychology |
| Goldie | White | Junior (3rd year) | Physics/Biology |

### B. Data Analysis

We engaged in collaborative coding, utilizing constant comparison with deductive coding, including both provisional and hypothesis coding [58,59]. Our theoretical model provided our initial a priori codes, and we also identified additional codes in our first cycle of analysis. The research team individually coded one interview and then met together to discuss and determine preliminary codes and their alignment with the model. Following the initial meeting, we coded the remainder of the interviews. Within this process, each interview was coded by at least two coders. Once a preliminary codebook was developed, it was tested. Testing was followed by a final analysis of the data. Once the data were coded, the final themes were discussed by the team. Dedoose was used to organize the data and to share the codes. In the paper below, we present the findings from the a priori codes identified in our theoretical model.

### IV. FINDINGS

In our analysis of students' open-ended responses to surveys and interview transcriptions, we looked for evidence of our hypothesized inputs to the development of student identity, including faculty and peer mentorship, career conceptualization, and leadership. In our full sample we hope to examine the differences in identity development between white and Latina women, but the current data did not allow for that comparison because of the complexities of institutional impact and intersectional identities.

### A. Mentorship

#### 1. Faculty Mentorship

The women in our study expressed that faculty mentors provided encouragement and helped them feel like members of a community. Several participants noted that their faculty mentors pushed them to try difficult courses, including Goldie, who stated, "[my professors] encouraged me to take … more challenging classes than I would have offered to take on my own". This encouragement helped these students feel like their faculty mentors saw them as capable, thus developing the students' sense of *competence* and making them feel *recognized* as a physics person. Goldie noted the importance of encouragement when trying courses and activities that she did not feel confident in: "So that was definitely a time at which [my faculty advisor] pushed me out of my comfort zone. I think I'm better for it, but for a while there, I was like, 'I don't know why I'm here. I'm going to drop this class.'" Goldie added, "Having professors that genuinely believe in your ability to succeed as a student, I think makes a huge difference." One of the ways mentors had a positive impact on the students' physics identity and *sense of belonging* was through this encouragement to try more challenging courses and activities, which provides recognition that helps build confidence in the students.

Gabriella, Violet, and Daria all noted that, at times, they felt they were struggling more than other physics students. Reassurance from faculty mentors that other students found the material challenging provided comfort and validation of their experience. Daria stated that her faculty mentor "is always telling me that, even when I don't feel like I'm doing a good job, that I'm doing fine and that there's people who are doing worse than me, which is really reassuring." This demonstrates how *recognition* from others of being a good physics student can provide validation. Similarly, Gabriella felt that "hearing my professor say, I know that research can be overwhelming and ... when he was a student, he also struggled. I think that really helped me to feel better." Similarly, Nina noted that hearing about the challenges her mentors faced as students contributed to her persisting in the program even when she had to retake multiple courses.

#### 2. Peer Mentorship

Our findings indicate that mentorship from peers was particularly important in helping the women engage with their physics community and made them feel *recognized* by their peers. Both Daria and Goldie started attending events at the encouragement of junior or senior women in the physics program. Daria says of her peer mentor, "If I go to events, I stand by her, which makes me feel like I'm not alone … otherwise, I just wouldn't go." Goldie echoes this sentiment by explaining how both her peer mentors "made a real effort to go out of their way to engage me in things. I'm the only woman in my class in physics, so both went out of their way to invite me to SPS events … and talk to me, and that made a difference." The peer mentors help foster a *sense of belonging* within the physics group, providing that *connection between social and physics identity* for many of these physics students.

Peer mentoring is also seen as a way to pay forward the help that the women have received. Goldie believes, "It's part of what makes our physics program so strong. That interclass relationship building because ... I'm not the only person that feels that way about what older students have done for me."

Similarly, Gabriella explained, "I really like to meet other girls who are interested in physics and help them go through that journey." These women *recognized* themselves as *belonging* in physics and were motivated to welcome other students to help others feel *like they belonged in physics*. The way peer mentors pulled these women into the social circles of the physics community was important for belonging in the same way that these same students then pulled in the next generation as part of their social identity.

Violet spoke about participating in the Women in STEM club and looking up to the club president because she made everyone feel heard and took everyone's opinion into account. Peer mentors provide women in physics with accountability and camaraderie. All participants discussed creating study groups, meeting regularly to do classwork, and providing each other with "emotional support," as stated by Violet. Additionally, Gabriella's peer mentor helped her prepare her graduate school applications, and Goldie found that doing lab research in close proximity with peers creates a bond. Working on physics problem-solving and lab research with their peers contributed to women's sense of *competence* and *performance.* Overall, the *sense of belonging* in a physics group was essential to the success of these students. All the women we interviewed gravitated toward other women as peer mentors.

## B. Career Conceptualization

We asked study participants about their future career plans, and how their experiences in physics influenced their decisions. Some students described plans to apply to graduate school and felt supported in this process. Other students perceived a difference between how pursuing graduate school versus other paths were viewed in their departments. Violet, who was considering graduate school in physics, noted a marked difference in how students were treated based on their career plans,

> Some of the professors really have an academic versus industry mindset. They will almost immediately [in] freshman year peg a student towards 'they're going to go to graduate school' or 'they're going to go to industry.' It's not necessarily due to their interests or even research ability. It's pretty [much] based off what their grades are going to look like… right off the bat, freshman year, kind of putting people into these categories, and in a way celebrating the students who are probably going to graduate school more than those who are going to go into industry, even though the majority of our students end up going into industry.

Similarly, Daria, who is headed to graduate school in a different field, stated:

> There's the people that the professors think will go to grad school and prepare them for grad school, And then there's the ones that they try to gear up for industry jobs. I think that the professors value the grad school section more than they value the industry people, even though you can be very successful in either one. I just think that they think higher of the people that they think should go to grad school.

These perceptions of how different career paths are treated are associated with complicated feelings around physics identity. Daria further explained, "I'm a math and physics double major. Whenever people ask me what my major is, I mainly just say math because there's a lot of connotations that come with being a physics major I don't want to associate myself with." Both Daria and Violet believed that their professors' perceptions of who was competent in physics played a role in how students were treated. Through the *recognition* of faculty and the implications for the development of feelings of *competence*, *career interests* made a difference for these students. What is less clear from these data, at least in part because many of the students were considering graduate school, was how knowing about career options outside of academia might influence students' sense of identity as a physics person.

## C. Leadership

The women in our study identified leadership roles they played in their physics organizations and noted the importance of leadership in developing their *confidence* as a physics person. Gabriella noted,

> I'm the secretary of Sigma Pi Sigma. Last semester, I helped everyone organize events and everything. And the president of Sigma Pi Sigma just asked me to be the vice president. … It really helped me to overcome that fear. … And it gives me more confidence about talking in public and interacting more with other people.

In addition to feeling more confident as a result of her leadership role, Nina also noted that she wanted to help other students feel more confident,

> As I've been in a leadership position, I've kind of been almost forced to present more confidence. Being a leader has increased my confidence, and I feel really great. I feel exceptionally grateful to, and really humbled to be in that position. And I really try my best to make sure that every single person that comes through…feels welcomed and people grow into, if they want physics that they become more confident in their physics. I try my best to do that.

For both Gabriella and Nina, serving in a leadership role helped them to feel like they belonged in physics, and felt *recognized* by others.

Leadership can be formal (associated with a position within an organization) or informal (leading without an organizational role). The leadership shown by the peer mentors in Section IV.A.2 often did not come with a formal title but was important to several of the women we interviewed. These peer mentorship roles were built on the student's *sense of belonging,* helped them share their *interest* with others, and positioned them to *recognize others* as being part of the community. Seeing women excelling in STEM and in leadership roles allows Violet to "look up to them in the way that it's kinda what I want to be when I grow up." In this way, peer mentors served as role models to other women in physics while also developing the mentors' own *sense of belonging* in physics. Our data do not yet give a good sense of how much these students see leadership, particularly informal leadership, as part of their identity and how much it builds their sense of *competence*, but the importance they place on helping the next group of students hints at its significance.

## V. DISCUSSION

In our analysis of surveys and interviews of undergraduate women in physics, we found that the mentorship of faculty and peers provided students with recognition of their competence in physics and encouraged engagement in the physics community. The push from faculty to take challenging courses made the students we talked to feel like their ability to do physics was recognized and helped build a sense of competence. In at least one case, a faculty mentor who noted his own struggles normalized the challenge and helped the student feel that struggle was not equated with a lack of competence. This supports previous literature on how positive mentoring practices play an important role in supporting the development of a physics identity and inoculating students against negative stereotypes and negative experiences that impact their self-efficacy and sense of belonging [49,50].

We also found that formal leadership experiences helped women in physics feel more confident, contributing to their recognition of themselves as physics people. Informal leadership experiences, including informal peer mentoring, lies at the intersection of social and physics identity. As has been noted in the literature, leadership and leadership identity are deeply entwined and built through social processes [43,44]. This framing of leadership fits well within our model, as we position leadership as another important part of a student's identity that overlaps with personal identity, social identity, and physics identity [5]. Nevertheless, it is unclear whether leadership identity stands as its own independent feature of identity for these students or whether leadership experiences serve more as inputs into recognition and competence, which, in turn, feed into a student's physics identity. While our preliminary results do not tease apart this distinction, formal and informal (particularly peer mentorship) experiences feed into these students' sense of belonging within the physics community as well as providing recognition of their abilities. In the future, we plan to look more deeply into the ways in which formal and informal leadership tie into physics students' sense of identity.

Being in the "academic career path" group either by choice or through faculty assessment provided several of these students with recognition of their physics abilities and supported their sense of competence in the field. Conversely, there was a negative impact on students' sense of competence and belonging for those who were not in this group. We sought to understand how knowing about career options outside of the academy supported women's physics identity, but our year-one findings did not offer insight beyond participants' beliefs that their physics professors viewed students differently based on their plans to go into academic research versus industry.

In these interviews, we did not find direct connections between mentorship, leadership, career conceptualization, and the building of interest or performance, but they did support the students' feelings of competence and recognition. We will continue to explore how these experiences contribute to the building of students' physics identity and how our model (see Fig. 1) can help explain the formation of physics identity.

For the three interview participants who were double majors, the other majors offered contrasting perspectives. Goldie described being the only woman in most of her physics classes but having women make up half of the biology majors. Even so, she said, "I get along with all the physics guys, but there's no denying that it's different fundamentally." Alternatively, Violet was more drawn to psychology because it seemed to have a better work-life balance. She found that in physics, she didn't like the "attitude towards other people just are not good towards their peers." Daria indicated that there were negative connotations with being a physics major and that it was a much more "cliquey" group, so she usually identified herself as a math major when asked. These different experiences don't provide direct information on identity development. Still, they do provide an interesting contrast to the students' experience in physics, and we will continue to explore these differences as we expand our sample.

This study has several limitations, including sample size and the number and types of undergraduate institutions represented by participants, which restricts our ability to generalize our findings beyond this study. It is clear from the interviews that the culture of the institution and department has played a significant role in the experiences of these women and their sense of identity and belonging, but more research is needed to understand if this applies more broadly in other institutional contexts. The sample size and participant institutional context also limit our ability to interpret and assess the usefulness of our theoretical model. While we found common themes among the women in our study, we are continuing to expand our sample as we expect that there is more to be learned from a wider range of student experiences. As we broaden our participant group, we are intentionally including students with a range of racial, ethnic, and cultural backgrounds to allow us to develop an understanding of the intersection of race and gender on the formation of physics identity. Another important limitation was that our interviews did not address some of the questions about career conceptualization and leadership, particularly informal leadership, deeply enough to fully illuminate how the students connected these things to feelings of recognition, competence, and performance. Because of the limits on our ability to connect career conceptualization and leadership to recognition, competence, and performance, their connections to self-efficacy and belonging were also limited. In our future research, we will investigate the potential relationship between gender and beliefs about careers in physics.

Our findings present a first look at how mentorship, leadership, and career conceptualization impact self-efficacy, belonging, and physics identity, framed primarily by the experiences of five undergraduate women in physics at two institutions. Understanding the interconnection between identity, belonging, and the support structures that help students persist in physics may help national organizations like SPS and Sigma Pi Sigma and their institutional-level student groups build effective programs for women and for all students in physics. As departments, clubs, and other physics groups look to develop programs to increase the persistence of women in physics, we want to deepen the understanding of how some of the key support activities relate to the development of identity.

## ACKNOWLEDGMENTS

Thank you to the students who have participated in this work. This work was supported under NSF award DUE-2044232.

[1] National Science Board, Science and Engineering Indicators 2020: The State of U.S. Science and Engineering, NSB-2020-1, National Science Foundation, 2020.
[2] *Newest Data Shows Mixed Progress for Women and Marginalized Groups in Physics Higher Ed*, http://www.aps.org/publications/apsnews/202211/newest-data.cfm.
[3] L. McNeil and P. Heron, *Preparing Physics Students for 21st-Century Careers*, Phys. Today **70**, 38 (2017).
[4] A. Y. Kim, G. M. Sinatra, and V. Seyranian, *Developing a STEM Identity Among Young Women: A Social Identity Perspective*, Rev. Educ. Res. **88**, 589 (2018).
[5] Z. Hazari, G. Sonnert, P. M. Sadler, and M.-C. Shanahan, *Connecting High School Physics Experiences, Outcome Expectations, Physics Identity, and Physics Career Choice: A Gender Study*, J. Res. Sci. Teach. n/a (2010).
[6] H. B. Carlone and A. Johnson, *Understanding the Science Experiences of Successful Women of Color: Science Identity as an Analytic Lens*, J. Res. Sci. Teach. **44**, 1187 (2007).
[7] J. P. Gee, *Chapter 3 : Identity as an Analytic Lens for Research in Education*, Rev. Res. Educ. **25**, 99 (2000).
[8] J. P. Gee, *Discourse Analysis: What Makes It Critical?*, in *An Introduction to Critical Discourse Analysis in Education.*, edited by R. Rogers (Routledge, 2004).
[9] E. Wenger, *Conceptual Tools for CoPs as Social Learning Systems: Boundaries, Identity, Trajectories and Participation*, in *Social Learning Systems and Communities of Practice*, edited by C. Blackmore (Springer London, London, 2010), pp. 125–143.
[10] J. D. Cribbs, Z. Hazari, G. Sonnert, and P. M. Sadler, *Establishing an Explanatory Model for Mathematics Identity*, Child Dev. **86**, 1048 (2015).
[11] R. Dou and H. Cian, *Constructing STEM Identity: An Expanded Structural Model for STEM Identity Research*, J. Res. Sci. Teach. **59**, 458 (2022).

[12] L. Espinosa, *Pipelines and Pathways: Women of Color in Undergraduate STEM Majors and the College Experiences That Contribute to Persistence*, Harv. Educ. Rev. **81**, 209 (2011).
[13] A. Godwin, G. Potvin, Z. Hazari, and R. Lock, *Identity, Critical Agency, and Engineering: An Affective Model for Predicting Engineering as a Career Choice*, J. Eng. Educ. **105**, 312 (2016).
[14] M. M. McDonald, V. Zeigler-Hill, J. K. Vrabel, and M. Escobar, *A Single-Item Measure for Assessing STEM Identity*, Front. Educ. **4**, 78 (2019).
[15] K. M. Paul, A. V. Maltese, and D. Svetina Valdivia, *Development and Validation of the Role Identity Surveys in Engineering (RIS-E) and STEM (RIS-STEM) for Elementary Students*, Int. J. STEM Educ. **7**, 45 (2020).
[16] J. M. Kane, *Young African American Children Constructing Academic and Disciplinary Identities in an Urban Science Classroom*, Sci. Educ. **96**, 457 (2012).
[17] A. T. Danielsson, H. King, S. Godec, and A.-S. Nyström, *The Identity Turn in Science Education Research: A Critical Review of Methodologies in a Consolidating Field*, Cult. Stud. Sci. Educ. **18**, 695 (2023).
[18] J. E. Stets, P. S. Brenner, P. J. Burke, and R. T. Serpe, *The Science Identity and Entering a Science Occupation*, Soc. Sci. Res. **64**, 1 (2017).
[19] M. M. Williams and C. E. George-Jackson, *USING AND DOING SCIENCE: GENDER, SELF-EFFICACY, AND SCIENCE IDENTITY OF UNDERGRADUATE STUDENTS IN STEM*, J. Women Minor. Sci. Eng. **20**, 99 (2014).
[20] E. Seymour and A.-B. Hunter, editors , *Talking about Leaving Revisited: Persistence, Relocation, and Loss in Undergraduate STEM Education* (Springer International Publishing, Cham, 2019).
[21] Elaine. Seymour and N. M. Hewitt, *Talking about Leaving : Why Undergraduates Leave the Sciences* (Westview Press, Boulder, Colo, 1997).
[22] G. Trujillo and K. D. Tanner, *Considering the Role of Affect in Learning: Monitoring Students' Self-Efficacy, Sense of Belonging, and Science Identity*, CBE—Life Sci. Educ. **13**, 6 (2014).
[23] L. Archer, E. Dawson, J. DeWitt, A. Seakins, and B. Wong, *"Science Capital": A Conceptual, Methodological, and Empirical Argument for Extending Bourdieusian Notions of Capital beyond the Arts*, J. Res. Sci. Teach. **52**, 922 (2015).
[24] G. Potvin et al., *Examining the Effect of Counternarratives about Physics on Women's Physics Career Intentions*, Phys. Rev. Phys. Educ. Res. **19**, 010126 (2023).
[25] A. Bandura, *Self-Efficacy: Toward a Unifying Theory of Behavioral Change*, (1977).
[26] A. Bandura, *The Explanatory and Predictive Scope of Self-Efficacy Theory*, J. Soc. Clin. Psychol. **4**, 359 (1986).
[27] D. H. Schunk and F. Pajares, *The Development of Academic Self-Efficacy*, in *Development of Achievement Motivation* (Elsevier, 2002), pp. 15–31.
[28] T. D. Fantz, T. J. Siller, and M. A. Demiranda, *Pre-Collegiate Factors Influencing the Self-Efficacy of Engineering Students*, J. Eng. Educ. Wash. DC **100**, 604 (2011).
[29] K. D. Multon, S. D. Brown, and R. W. Lent, *Relation of Self-Efficacy Beliefs to Academic Outcomes: A Meta-Analytic Investigation.*, J. Couns. Psychol. **38**, 30 (1991).
[30] T. Luo, W. W. M. So, Z. H. Wan, and W. C. Li, *STEM Stereotypes Predict Students' STEM Career Interest via Self-Efficacy and Outcome Expectations*, Int. J. STEM Educ. **8**, 36 (2021).
[31] H.-B. Sheu, R. W. Lent, M. J. Miller, L. T. Penn, M. E. Cusick, and N. N. Truong, *Sources of Self-Efficacy and Outcome Expectations in Science, Technology, Engineering, and Mathematics Domains: A Meta-Analysis*, J. Vocat. Behav. **109**, 118 (2018).
[32] A. M. Zaniewski and D. Reinholz, *Increasing STEM Success: A near-Peer Mentoring Program in the Physical Sciences*, Int. J. STEM Educ. **3**, 14 (2016).

[33] S. Cwik and C. Singh, *Damage Caused by Societal Stereotypes: Women Have Lower Physics Self-Efficacy Controlling for Grade Even in Courses in Which They Outnumber Men*, Phys. Rev. Phys. Educ. Res. **17**, 020138 (2021).

[34] S. Banchefsky, K. L. Lewis, and T. A. Ito, *The Role of Social and Ability Belonging in Men's and Women's pSTEM Persistence*, Front. Psychol. **10**, 2386 (2019).

[35] Z. Y. Kalendar, E. Marshman, C. Schunn, T. Nokes-Malach, and C. Singh, *Why Female Science, Technology, Engineering, and Mathematics Majors Do Not Identify with Physics: They Do Not Think Others See Them That Way.*, Phys. Rev. Phys. Educ. Res. **15**, (2019).

[36] L. Avraamidou, *Identities in/out of Physics and the Politics of Recognition*, J. Res. Sci. Teach. **59**, 58 (2022).

[37] N. J. Hiller, An Examination of Leadership Beliefs and Leadership Self-Identity: Constructs, Correlates, and Outcomes, Ph.D., The Pennsylvania State University, 2005.

[38] M. J. McCormick, J. Tanguma, and A. S. López-Forment, *Extending Self-Efficacy Theory to Leadership: A Review and Empirical Test*, J. Leadersh. Educ. **1**, (2002).

[39] M. L. Gallagher, J. C. Marshall, M. L. Pories, and M. Daughety, *Factors Effecting Undergraduate Leadership Behaviors*, J. Leadersh. Educ. **13**, 46 (2014).

[40] A. C. Maia, Educating Culturally Relevant Leaders: Experiences in Leadership Identity, Capacity, and Efficacy Development in College Students, ProQuest Dissertations Publishing, 2022.

[41] J. Blaney, *UNDERGRADUATE STEM LEADERSHIP: UNDERSTANDING THE GENDER GAP IN SELF-RATED LEADERSHIP ABILITY BY EXPLORING WOMEN'S MEANING-MAKING*, J. Women Minor. Sci. Eng. **26**, 177 (2020).

[42] Close, Eleanor, *Becoming Physics People: Development of Integrated Physics Identity through the Learning Assistant Experience*, (n.d.).

[43] D. DeRue, S. Ashford, and N. Cotton, *Assuming the Mantle: Unpacking the Process by Which Individuals Internalize a Leader Identity*, Explor. Posit. Identities Organ. Build. Theor. Res. Found. (2009).

[44] D. S. Derue and S. J. Ashford, *WHO WILL LEAD AND WHO WILL FOLLOW? A SOCIAL PROCESS OF LEADERSHIP IDENTITY CONSTRUCTION IN ORGANIZATIONS*, Acad. Manage. Rev. (2010).

[45] K.-Y. Chan and F. Drasgow, *Toward a Theory of Individual Differences and Leadership: Understanding the Motivation to Lead.*, J. Appl. Psychol. **86**, 481 (2001).

[46] R. Kark and D. Van Dijk, *Motivation to Lead, Motivation to Follow: The Role of the Self-Regulatory Focus in Leadership Processes*, Acad. Manage. Rev. **32**, 500 (2007).

[47] T. C. Dennehy and N. Dasgupta, *Female Peer Mentors Early in College Increase Women's Positive Academic Experiences and Retention in Engineering*, Proc. Natl. Acad. Sci. **114**, 5964 (2017).

[48] Board on Higher Education and Workforce, Policy and Global Affairs, and National Academies of Sciences, Engineering, and Medicine, *Effective Mentoring in STEMM: Practice, Research, and Future Directions: Proceedings of a Workshop—in Brief* (National Academies Press, Washington, D.C., 2017).

[49] N. Dasgupta, *Ingroup Experts and Peers as Social Vaccines Who Inoculate the Self-Concept: The Stereotype Inoculation Model*, Psychol. Inq. **22**, 231 (2011).

[50] J. G. Stout, N. Dasgupta, M. Hunsinger, and M. A. McManus, *STEMing the Tide: Using Ingroup Experts to Inoculate Women's Self-Concept in Science, Technology, Engineering, and Mathematics (STEM).*, J. Pers. Soc. Psychol. **100**, 255 (2011).

[51] Committee on Effective Mentoring in STEMM, Board on Higher Education and Workforce, Policy and Global Affairs, and National Academies of Sciences, Engineering, and Medicine,

[51] *The Science of Effective Mentorship in STEMM* (National Academies Press, Washington, D.C., 2019).
[52] M. Ong, *Body Projects of Young Women of Color in Physics: Intersections of Gender, Race, and Science*, Soc. Probl. **52**, 593 (2005).
[53] *Physics Bachelors 1 Year Later*, https://www.aip.org/statistics/physics-trends/physics-bachelors-1-year-later-1.
[54] *First-Year Graduate Students in Physics and Astronomy: Characteristics and Background | American Institute of Physics*, https://www.aip.org/statistics/reports/first-year-graduate-students-physics-and-astronomy-characteristics-and.
[55] H. Sauermann and M. Roach, *Science PhD Career Preferences: Levels, Changes, and Advisor Encouragement*, PLoS ONE **7**, e36307 (2012).
[56] Z. Hazari, D. Chari, G. Potvin, and E. Brewe, *The Context Dependence of Physics Identity: Examining the Role of Performance/Competence, Recognition, Interest, and Sense of Belonging for Lower and Upper Female Physics Undergraduates*, J. Res. Sci. Teach. **57**, 1583 (2020).
[57] D. A. Dillman, L. M. Christian, and J. D. Smyth, *Internet, Phone, Mail, and Mixed-Mode Surveys : The Tailored Design Method*, 4th edition. (Wiley, Hoboken, 2014).
[58] Ph. D. Barney G. Glaser, *The Constant Comparative Method of Qualitative Analysis*, Grounded Theory Rev. **7**, (2008).
[59] J. Saldaña, *The Coding Manual for Qualitative Researchers*, Fourth edition. (SAGE Publications, London, 2021).